\documentclass{bmcart}

\usepackage{soul}
\usepackage{amsthm,amsmath}
\usepackage{amssymb}
\usepackage{cite}
\usepackage{todonotes}
\usepackage{graphicx}
\graphicspath{{figures/}}
\RequirePackage{hyperref}
\usepackage[utf8]{inputenc}

\begin{document}

\begin{frontmatter}

\begin{fmbox}
\dochead{Research}

\title{Acoustic Cloak Design via Machine Learning}

\author[
  addressref={aff1}, 
  email={thang.tran01@sjsu.edu} 
]{\inits{T.T.}\fnm{Thang} \snm{Tran}}
\author[
  addressref={aff2},
   corref={aff2},  
  email={feruza.amirkulova@sjsu.edu}
]{\inits{F.A.}\fnm{Feruza} \snm{Amirkulova}}
\author[
  addressref={aff1},
  email={ehsan.khatami@sjsu.edu}
]{\inits{E.K.}\fnm{Ehsan} \snm{Khatami}}

\address[id=aff1]{%
  \orgdiv{Department of Physics and Astronomy}, 
  \orgname{San Jose State University},          
  \city{San Jose},                            
  \cny{USA}                                   
}
\address[id=aff2]{%
  \orgdiv{Department of Mechanical Engineer},
  \orgname{San Jose State University},
  \city{San Jose},
  \cny{USA}
}

\end{fmbox}

\begin{abstractbox}

\begin{abstract}  
Acoustic metamaterials are engineered microstructures with special mechanical and acoustic properties enabling exotic effects such as wave steering, focusing and cloaking. The design of acoustic cloaks using scattering cancellation has traditionally involved the optimization of  metamaterial structure based on direct computer simulations of the total scattering cross section (TSCS) for a large number of configurations. Here, we work with sets of cylindrical objects confined in a region of space and use machine learning methods to streamline the design of 2D configurations of scatterers with minimal TSCS demonstrating cloaking effect at discrete sets of wavenumbers. After establishing that artificial neural networks are capable of learning the TSCS based on the location of cylinders, we develop an inverse design algorithm, combining variational autoencoders and the Gaussian process, for predicting optimal arrangements of scatterers given the TSCS. We show results for up to eight cylinders and discuss the efficiency and other advantages of the machine learning approach. 
\end{abstract}


\end{abstractbox}

\end{frontmatter}

\section*{Introduction}
Acoustic metamaterials are engineered microstructures with special mechanical and acoustic properties 
 such as negative effective mass density, bulk modulus, and refractive index \cite{Kadic13b,Mejica13,Su2016,Titovich2016}. 
 The unique arrangement of components in metamaterials gives rise to exotic interactions with acoustic 
 waves at large wavelengths. This has many applications such as wave steering \cite{Packo2021}, 
 cloaking \cite{Norris08b,Amirkulova2020}, and focusing \cite{TitovichNorrisHabberman2016, Su2016,Fahey2019}. 
 The traditional method of designing acoustic cloaking devices involves direct computer simulation~\cite{Haberman_AM13}, 
 topology optimization~\cite{AndkjAer2013}, stochastic optimization~\cite{HaSkansson2007,Sanchez-Dehesa2011}, 
 and gradient based optimization\cite{Amirkulova2020, Andersen19} of the acoustic response based on the structure.
The direct forward methods require the iterative process of trial and error, leading to sub-optimal performance 
of metadevices. The computational costs for non-convex optimization problem in a high-dimensional space 
including stochastic optimization, adjoint based design and gradient based optimization are high and 
present an immense computational challenge~\cite{CampbellFan2019ReviewOptimizMeta,Elsawy20}. While these 
approaches have shown improvement in efficiency and performance over the years, they are still ineffective 
in the inverse design of broadband metamaterials and metadevices, specifically in predicting the structure 
parameters given the acoustic response at different values of frequencies or incident angles.

The last decade has witnessed a surge of scientific publications in which deep learning, reinforcement 
learning and generative modelling were applied in different areas of science and 
engineering~\cite{Barrett2019,Elton2019,Tahersima2019,So2020}. Recent advances in the field of machine 
learning have enabled a new, data driven, approach with a great promise to solve problems such as the 
inverse design in acoustic metamaterials \cite{Gurbuz21,Shah_et_al_Amirkulova21,Wu2021b,Wu2022,Ahmed2021}. 
Early machine learning applications in acoustic forward design date back to the late 90s when Jenison 
first used spherical basis function of fully-connected neural networks (FC) for approximating the acoustic 
scattering of a rigid scatterer \cite{Jenison86,Jenison98}. Hesham and El-Gamal \cite{Hesham08,Hesham08a} 
later solved an integral equation of acoustic scattering using wavelet basis and FC.

More recent works have proven that machine learning is much more efficient than numerical methods in 
both forward and inverse design of metamaterials. For example, Gnecco et al. \cite{Gnecco2021} employed 
principal component analysis (PCA) to gradient fields in spectral design of acoustic metamaterials, 
solving constrained nonlinear optimization problems. Fan et al.\cite{Fan2020} formulated acoustic 
scattering by a single scatterer as a 2D image-to-image regression problem using convolutional neural 
networks (CNN). In their model, the inputs were the images of convex prism objects, including circular 
and ellipsoidal cylinders,  and square bars; the outputs were the loudness fields computed using the 
Triton system that employs the fast adaptive rectangular decomposition pseudo-spectral wave solver. 
Fan et al.\cite{Fan2020c} developed a CNN to predict the object geometry from acoustic scattering 
given the images of  total acoustic field as inputs to CNN. Meng et al. \cite{Meng2020} investigated 
the inverse acoustic scattering problem that reconstructs the obstacle shape with far-field information 
using fully connected neural network (FC). Shah et al.\cite{Shah_et_al_Amirkulova21} employed deep 
reinforcement learning algorithms to design acoustic cloaks by adjusting positions and radii of 
cylindrical structures; these reinforcement learning models are capable of predicting better results 
than the-state-of-the-art gradient based optimization algorithms. Wu et al. \cite{Wu2021b} proposed 
machine learning  framework for the design of one-dimensional periodic and non-periodic acoustic 
metamaterials using deep learning and reinforcement learning algorithms. Alternative machine learning 
methods such as powerful deep generative models, including generative adversarial networks~\cite{Goodfellow14,Gurbuz21}, 
autoencoders \cite{ Ahmed2021,Wu2022}, and variational autoencoders (VAE) \cite{Ahmed2021,Kingma2014,Ma2019} 
are also able to produce synthetic structures after being trained on real examples.

Most of the studies mentioned above succeed with predicting the acoustic response based on the structures. 
However, the challenge of finding a streamlined approach in designing structures with a desired acoustic 
response is still in its infancy. In this study, we continue tackling this inverse design problem by proposing 
a promising new method. First we show that FC and convolutional neural networks can be trained to predict 
the acoustic response based on the locations of cylindrical structures (forward design). We then implement 
a VAE to convert the structures into a latent space and train it together with a regressor that is capable 
of yielding the total scattering cross section (TCSC) given the latent variables. By performing Bayesian 
optimization using the Gaussian process (GP) in the continuous latent space, we demonstrate that one can 
quickly find new configurations with minimal TSCS. Our method, which is straightforward to implement and 
computationally cheap to operate after training, has great implications for metamaterials design with 
on-demand properties.

\section*{Method}

\subsection*{Multiple Scattering Theory}

In order to analyze how sound is scattered in acoustic media, we use the multiple scattering 
theory \cite{Martin06}. We consider multiple scattering in the context of the acoustic time harmonic 
wave equation in two dimensions. The total acoustic pressure field $p({\bf x})$, ${\bf x}\in {\cal R}^2$
is defined as the sum of incident $p_{inc}$ and scattered $p_{sc}$ pressure fields:
\begin{equation}\label{eq1}
p = p_{inc} + p_{sc},
\end{equation}
which satisfies the Helmholtz  equation:
\begin{equation}\label{eq2}
\nabla^2 p + k^2 p = 0,
\end{equation}
where $ k = \omega /c$ is the wavenumber, $c$ is the acoustic speed, and $\omega$ is the frequency. 
As a measure of the scattering, we use the TSCS which is denoted by $\sigma$. We use the optical 
theorem \cite{Norris2015} to relate the TSCS to the  scattering   amplitude in the forward direction, 
i.e. the direction of propagation of the incident plane wave, here assumed to be  the positive 
$x-$direction \cite{Amirkulova2020}:
\begin{equation}\label{eq3}
 \sigma = -2\textrm{Re} f(0), 
\end{equation}
where the far-field amplitude form function, $f = f(\theta,\, {\bf r}_1, \, {\bf r}_2, \ldots, {\bf r}_M)$, 
$\theta=\arg({\bf x} )$, is the angular part of the scattered pressure $p_{sc}$ in the far-field in 
terms of position vector, and ${\bf r}_1, \, {\bf r}_2, \ldots, {\bf r}_M$ denote  the positions of 
each scatterer. Here, $f(0)=f(\theta=0)$. A more detailed description of multiple scattering problem 
formulation can be found in Refs. \cite{Martin06,Shah_et_al_Amirkulova21,Amirkulova2020, AmirkulovaPhD2014}.
{\em Our goal in this study is to find an efficient way to obtain scatterer locations that minimize 
$\sigma$ at certain wavenumbers (inverse design), and hence, pave the way for the creation of acoustic cloaks.}

\subsection*{Data Generation}

\begin{figure}[h!]
    \centering\includegraphics[width=0.97\linewidth]{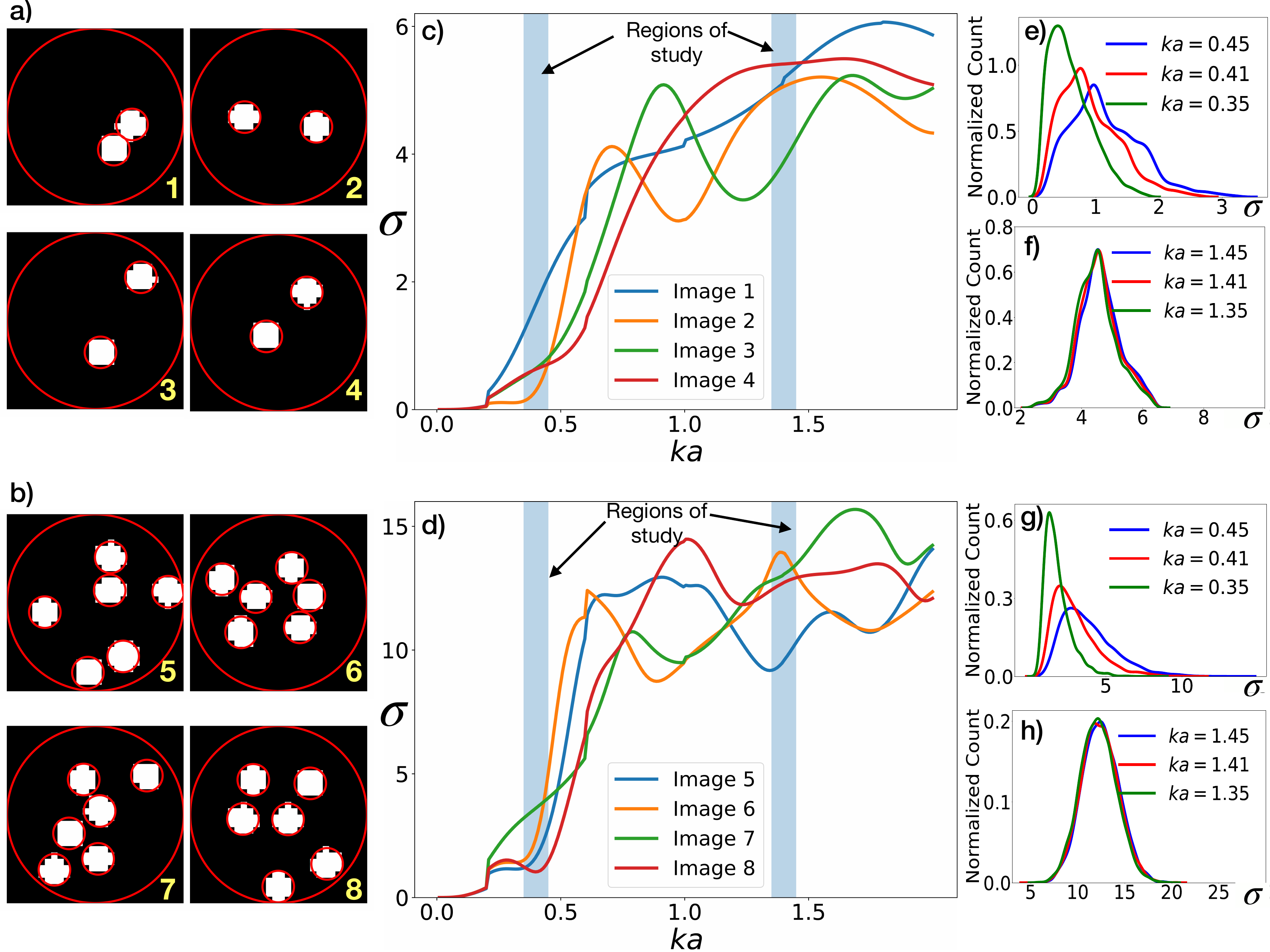}
    \caption{Random configurations generated in binary images, (a) and (b), and their corresponding 
    total scattering cross section $\sigma$. (c) and (d) show the TSCS values as a function of $ka$ 
    for $4$ different $28\times28$ images of $M=2$ and $M=6$ from panel (a) and (b). The white dots 
    represent the rigid cylindrical scatterers of radii $a$. The scatterers in (a) and (b) are 
    confined to an artificial circle of radius $R=6a$ with minimum distance between the cylinder 
    centers $|{\bf r}_i- {\bf r}_j|= 2a+\delta$, where $\delta = 0.1a$. (e)-(h) show the normalized 
    count of the total 
    scattering cross section used for training. The three sample wavenumbers shown in each plot 
    belong to two different regions, $k a \in [0.35, 0.45]$ and $k a \in [1.35, 1.45]$. In the smaller 
    $ka$ region, distributions gradually shift to the right for larger wavenumbers while they mostly 
    overlap in the larger $ka$ region.}
    \label{fig1}
\end{figure}

Datasets for TSCS of different configurations of rigid cylinders are generated using the multiple 
scattering solver \cite{Amirkulova2020, AmirkulovaPhD2014} implemented in MATLAB. For a given 
configuration of scatterers, $\sigma$ is evaluated at discrete values of the normalized wavenumber $ka$. 
We first randomly position $M$ uniform cylinders of radius $a$ inside an artificial circular region of radius 
$R=6a$ as depicted in Fig.~\ref{fig1} (a) and (b). We keep the minimum distance between the 
cylinder centers as $|{\bf r}_i-  {\bf r}_j|= 2a+\delta$ where  $\delta = 0.1a$ \cite{Amirkulova2020}. 
The images contain $28 \times 28$ binary pixels; 
taking the value $1$ for the area of rigid cylinders, denoted by the color white, and $0$ for the 
exterior fluid, denoted by the color black. Then $\sigma(k_ia)$ is evaluated at 11 discrete values of 
wavenumber $ka \in [0.35,0.45]$ and $ka \in [1.35, 1.45]$ using Eq.~\eqref{eq3}. Fig.~\ref{fig1}(a)-(d) 
show sample configurations for $M=2$ and $M=6$ along with their corresponding $\sigma$'s as functions 
of the wavenumber. Figs.~\ref{fig1}(e)-(h) show histograms of $\sigma$ over the entire data at select 
$ka$'s within two different intervals that we have considered in this study.

\subsection*{Forward Design}

We employ both FC and CNN in our forward design. The goal is for the feed-forward artificial neural 
network to predict the TSCS (learn the nonlinear function of $ka$) given the scatterer positions. We provide 
the networks with  binary images of scatterer configuration as input for the training. The dataset for 
each $M$ is split into $54,000$ ($90$\%) samples for training and $6,000$ ($10$\%) samples for unbiased 
validation during training. Experimenting with different image sizes ranging from a few hundreds to 
thousands of pixels, we find that the image resolution does not affect the outcome significantly, and 
therefore, we have chosen $28\times28$ images for training to avoid unnecessary computational costs.

For the FC, we design a 3-hidden-layer network with 100 nodes each. In that case, the $28\times28$ input 
image is flattened into a one-dimensional vector of length $784$ before feeding it into the fully connected 
layers. Since all of the pixel values are used for training, the neural network is highly susceptible to 
overfitting when training with $M\ge 6$. Hence, a dropout layer with rate of $0.2$ is added after the 
third hidden layer.

\begin{figure}[h!]
    \centering\includegraphics[width=0.96\linewidth]{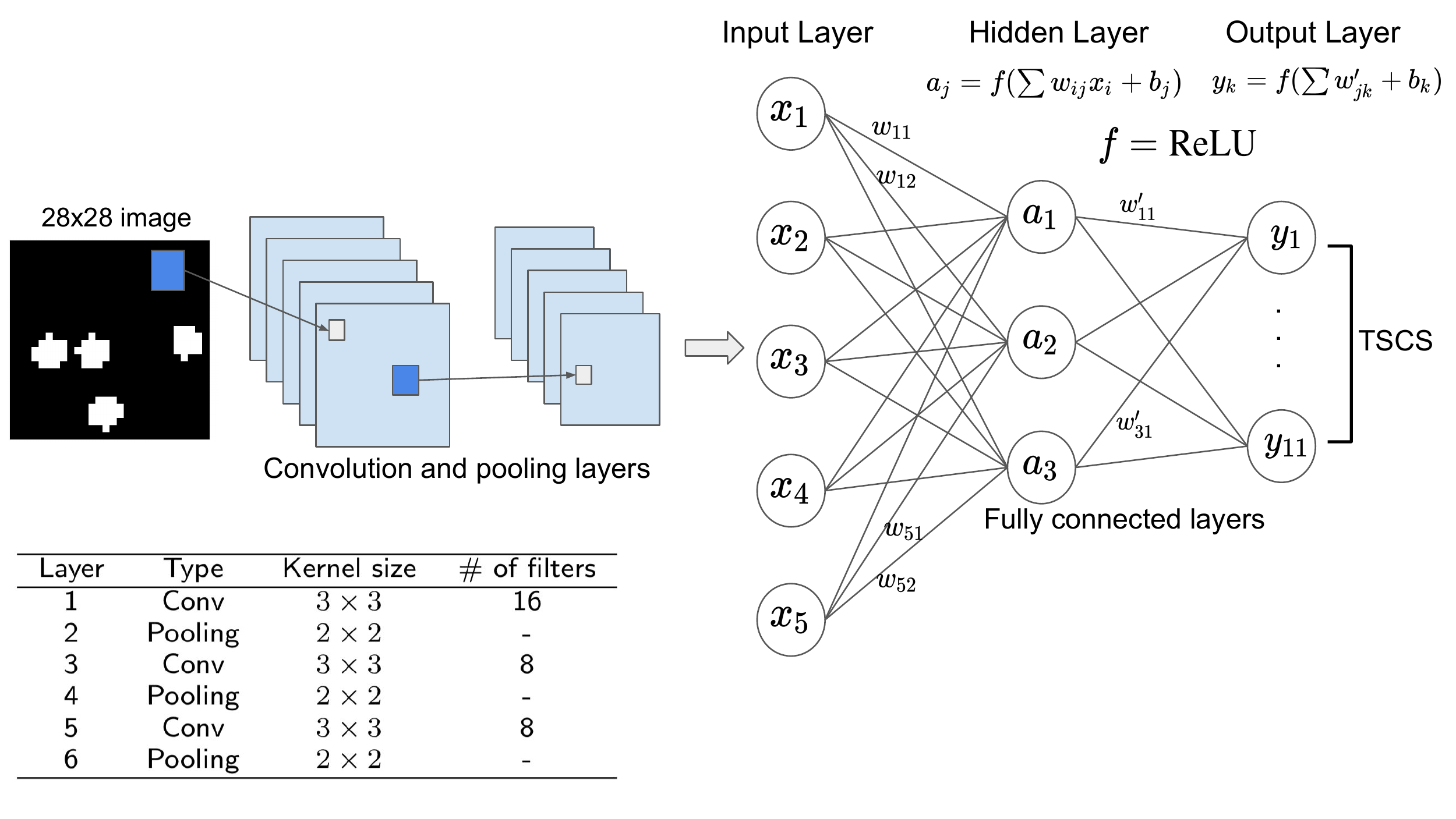}
    \caption{A diagram of forward design used for TSCS regression. The input image is first passed through 
    a series of convolutional and pooling layers in which the kernel, depicted by the small blue square, 
    sweeps through the input image to extract features and reduce the dimensions. The table below shows the 
    details of each layer. Next, the resulting output is flattened and fed into a fully connected neural 
    network. The input layer $x_i$ no longer represents the pixel values from the original image but rather 
    abstract image attributes. The activation function of perceptrons is $f =$ ReLU (rectified linear unit). 
    The output layer comprising of 11 perceptrons for the TSCS values, which are compared to the true values 
    during training by the loss function of mean squared error. }
    \label{fig2}
\end{figure}

The CNN is our primary focus because of its ability to leverage spatial correlations in an image, which 
would otherwise be lost during image flattening step of a FC. Fig.~\ref{fig2} illustrates an example of 
our CNN architecture. In contrast to the FC, the CNN passes an image through a series of convolution and 
pooling layers for feature extraction. Specifically, in the first convolution layer a $3\times3$ kernel 
matrix slides across the image, convolving with portions of it, and passing the information to later layers. 
In the first pooling layer, only maximum pixel values in $2\times2$ blocks of feature maps are selected, 
reducing the resolution and removing redundant information in order to mitigate overfitting. Finally, the 
output of the last pooling layer is flattened and fed into one layer of fully connected nodes before 
decision making in the output layer. Overall, the network learns to make prediction by optimizing its 
weights and biases. These parameters are continually adjusted during the training process until the mean 
squared error between its output and the exact TSCS saturates (see Appendix).

\subsection*{Inverse Design}

\begin{figure}[htbp]
    \centering\includegraphics[width=0.96\linewidth]{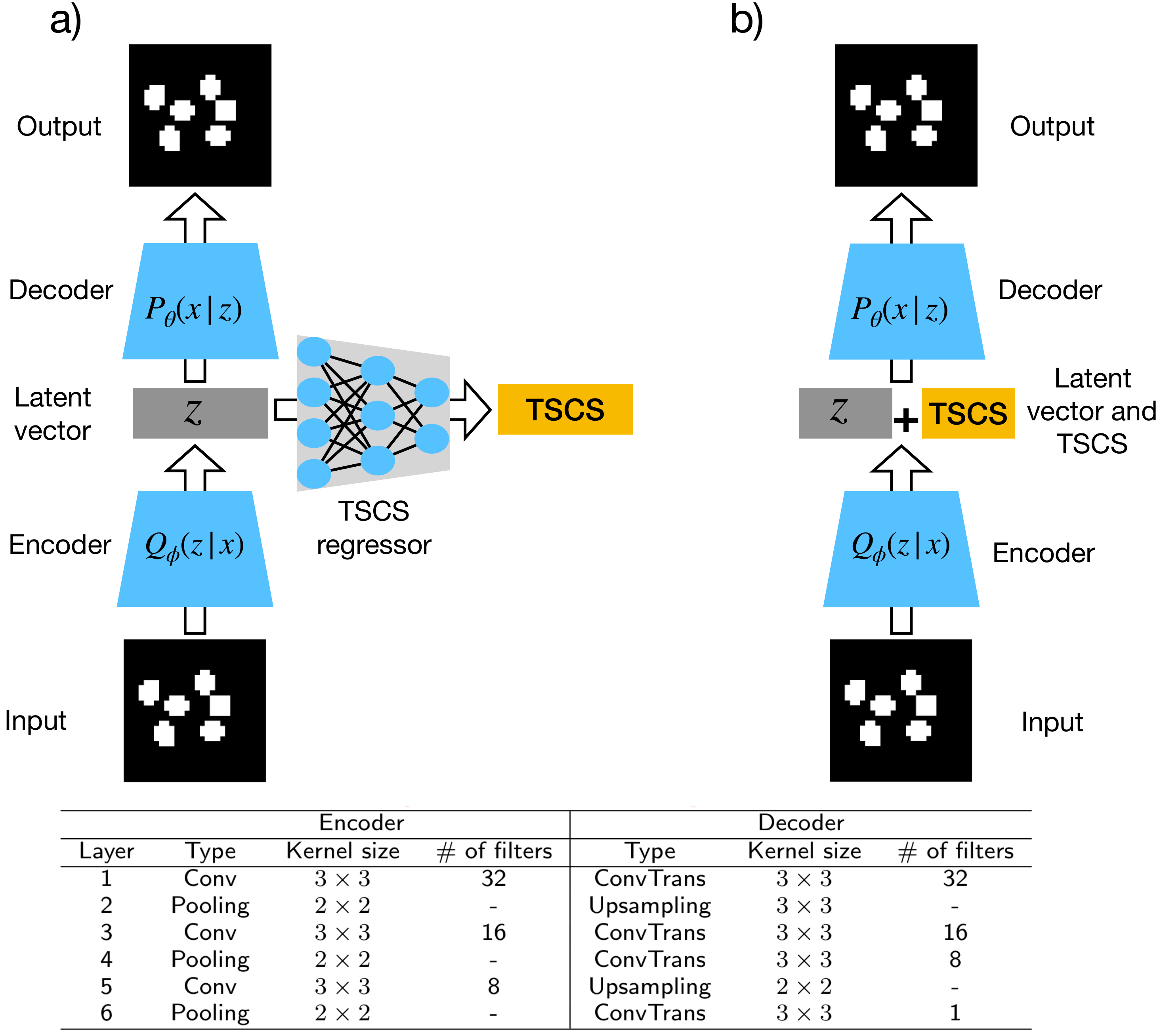}
    \caption{The two variational autoencoder (VAE) architectures used for inverse design: a) supervised 
    VAE  and b) conditional VAE. Starting from an image of scatterers, the encoder transforms the image 
    into a lower-dimensional, continuous representation vector $z$. In a) this latent vector is fed into 
    a TSCS neural network regressor made of 3 layers, whereas in b) the TSCS is directly embedded into 
    the latent by concatenating the two vectors. The decoder converts the latent vector back into the 
    original input image. By incorporating the TSCS into VAEs, we condition the networks to organize the 
    latent variables in accordance with the physical information offered by the TSCS. We then use this 
    latent space to perform optimization and search for the configuration with the lowest TSCS. The table 
    at the bottom shows the layers used in the encoder and the decoder. We used ReLU as the activation 
    for the middle layers and sigmoid for the output layer. (Conv: convolution; ConvTrans: convolution 
    transpose)}
    \label{fig3}
\end{figure}

VAE is a probabilistic generative model that uses Bayesian inference combined with a neural network to 
approximate the distribution of data \cite{Kingma14}. The model learns from examples and compresses the 
original data into a low dimensional and continuous representation called latent space, characterized by 
a latent vector $z = (z_1, z_2, \dots)$. The number of latent variables generally determines the amount 
of information or relevant features that can be encoded into the latent space. We apply VAE to our datasets 
of images of different cylindrical configurations. After training the model, sampling from the latent 
space allows us to generate new, unseen images. However, in addition to the plain VAE, which simply 
approximates the input image data, we {\it condition} the VAE with TSCS, so that the latent space does 
not only reflect the positional characteristics of the scatterers but also their corresponding TSCS. 
We perform this conditioning procedure within two different methods: supervised VAE (SVAE) \cite{SVAE}
and conditional VAE (CVAE) \cite{CVAE}. We then examine the statistics of the latent variables from the 
two modified VAE models and search for new configurations with minimum TSCS. Figure~\ref{fig3} shows the 
details of the VAEs architecture and layers used to train the models in our study.

The goal of VAEs is to estimate the true distribution of input images $P_\theta(x)$ using a neural network:
\begin{equation}\label{truePx}
P_\theta(x) = \int P_\theta(x| z)P(z)dz =\int P_\theta(z| x)P(x)dz, 
\end{equation}
where $x$ is the input data, $z$ is the latent variable drawn from a predefined distribution $P(z)$, 
and $\theta$ are the model parameters. The problem with maximizing the likelihood of obtaining $z$ given 
$x$ is that the above integral is intractable due to the large number of parameters $\theta$ produced by 
the neural network. Instead, VAEs approximate $P_\theta(z | x)$ with an encoder $Q_\phi(z | x)$, which 
is often chosen to be a multivariate Gaussian distribution: 
\begin{equation}\label{encoderQ}
Q_\phi(z | x) = \mathcal{N}(z;\mu(x), diag(s(x)))
\end{equation}
The above equation implies that each latent variable, after being trained on data, should follow a 
normal distribution with mean $\mu(x)$ and standard deviation $s(x)$, independently of one another. 
The inference model $Q_\phi$ takes in the input image and generates the latent variable $z$ while the 
decoder $P_\theta(x | z)$ reconstruct the original input based on $z$, as depicted in Fig. \ref{fig3}. 
The difference between the two conditional distributions is determined by the Kullback-Leibler (KL) divergence: 
\begin{equation}\label{DKL}
D_{KL}(Q_\phi(z | x) | P_\theta(z | x)) = \mathbb{E}_{z \sim Q}[\log Q_\phi(z | x) - \log P_\theta(z | x)].
\end{equation}
Using Bayes theorem, $P_\theta(z | x) = P_\theta(x|z)P(z)/P_\theta(x)$,
we replace $P_\theta(z | x)$ in Eq.~\eqref{DKL} to obtain
\begin{equation}\label{DKLfinal}
\log P_\theta(x) - D_{KL}(Q_\phi(z | x) | P_\theta(z | x)) = \mathbb{E}_{z \sim Q}[\log P_\theta(x | z)] 
- D_{KL}(Q_\phi(z | x) | P(z))
\end{equation}
Maximizing the left side of Eq.~\eqref{DKLfinal}, also known as finding the variational lower bound 
or evidence lower bound, means optimizing the model parameters $\phi$ and $\theta$ of the neural 
network via backpropagation. As the model is trained, the encoder $Q_\phi(z | x)$ and the decoder 
$P_\theta(x | z)$ will become better at encoding the attributes $z$ given data $x$ and reconstructing 
the data $x$ given the latent vector $z$. In other words, we want to minimize the KL divergence on the 
left hand side of Eq.~\eqref{DKLfinal}, which translates to maximizing the term 
$\mathbb{E}_{z \sim Q}[\log P_\theta(x | z)]$, or minimizing the reconstruction loss ($\mathcal{L}_R$), 
and minimizing the KL divergence loss ($\mathcal{L}_{KL}$) in the right hand side of Eq.~\eqref{DKLfinal}. 
The former is simply the mean squared error between the input and the reconstructed image, and the 
latter is accomplished by adjusting the network parameters in the encoder ($\phi$) during training to 
match the network distribution for $z$ given the images and the desired normal distribution.

This training of the VAE requires reparameterization of the encoded sample $z$ drawn  from the 
predefined normal distribution as: 
\begin{equation}\label{epsison}
z = \mu + \epsilon s
\end{equation}
where $\epsilon \sim \mathcal{N}(0,1)$ \cite{Kingma14}. This ensures backpropagation gradients 
passing through all layers.

In the next step, in order to implement a SVAE and bring in the physical property of interest, we add 
another neural network regression model that takes in the encoded latent variables $z$ as input and 
predicts the TSCS. The TSCS regression model loss ($\mathcal{L}_{TSCS}$) is minimized simultaneously 
along with KL and reconstruction loss during training:
\begin{equation}\label{losses}
\mathcal{L}_{SVAE} = \mathcal{L}_R + \mathcal{L}_{KL} + \mathcal{L}_{TSCS}
\end{equation}

In addition, we also experimented with CVAE by directly embedding the TSCS into the input data used 
for training. In that case, Eq.~\eqref{DKLfinal} becomes: 
\begin{equation}\label{DKLc}
\begin{aligned}[b]
& \log P_\theta(x|c) - D_{KL}(Q_\phi(z | x,c) | P_\theta(z | x,c))\\
&\quad = \mathbb{E}_{z \sim Q}[\log P_\theta(x | z,c)] - D_{KL}(Q_\phi(z | x,c) | P_\theta (z|c)),
\end{aligned}
\end{equation}
where $c$ is the {\em conditional vector}, in our case the TSCS corresponding to each input image, that 
we concatenate with $z$. The TSCS regressor, which was used as the objective function to minimize in the 
SVAE, is substituted in this case by the combination of the decoder part of the VAE and the trained CNN 
model from forward design, which takes as input the images produced by the decoder.

After acquiring a continuous representation of the data through the latent variable, we perform global 
optimization in the latent space using a Gaussian process \cite{Osborne09gaussianprocesses}. The Gaussian 
process is extremely robust in predicting any smooth and nonlinear function based on data. First, we 
obtain a surrogate function by training a GP model to approximate the TSCS regressor given the latent 
variables as input. Then, we minimize this surrogate function through Bayesian optimization to search for 
$z$'s that correspond to configurations with the lowest TSCS. In the next step, we decode those $z$'s 
back to images using the trained decoder from our VAE models. Finally, the suggested positions of the 
cylinders in the decoded image are converted back into physical coordinates to compute the TSCS with our 
analytical solution for comparison.

Optimizing the probabilistic GP model, as opposed to working with the trained fully connected neural 
network regressor, is the key step here that allows us to perform the inverse design. Moreover, it 
provides a smoother search space and the function is cheaper to evaluate too. The advantage of including 
the regressor in the SVAE, or concatenating $z$ with TSCS in the CVAE, is that we are conditioning the 
latent variables to reflect TSCS, and hence, forcing the neural network model to encode the physics of 
the problem as opposed to simply the whereabouts of the scatterers.

\section*{Results}

We used Tensorflow \cite{Abadi15} python libraries with Keras application programming interface, the 
open source machine learning framework, to build, train, and test our models. We first discuss the 
performance and limitations of FC and CNN in forward design. Next, we report the outcome of VAEs in 
inverse design followed by the optimization results of the TSCS with respect to the latent variable.

\subsection*{Convolutional Neural Networks and the Forward Design}

\begin{figure}[h!]
    \centering\includegraphics[width=0.97\linewidth]{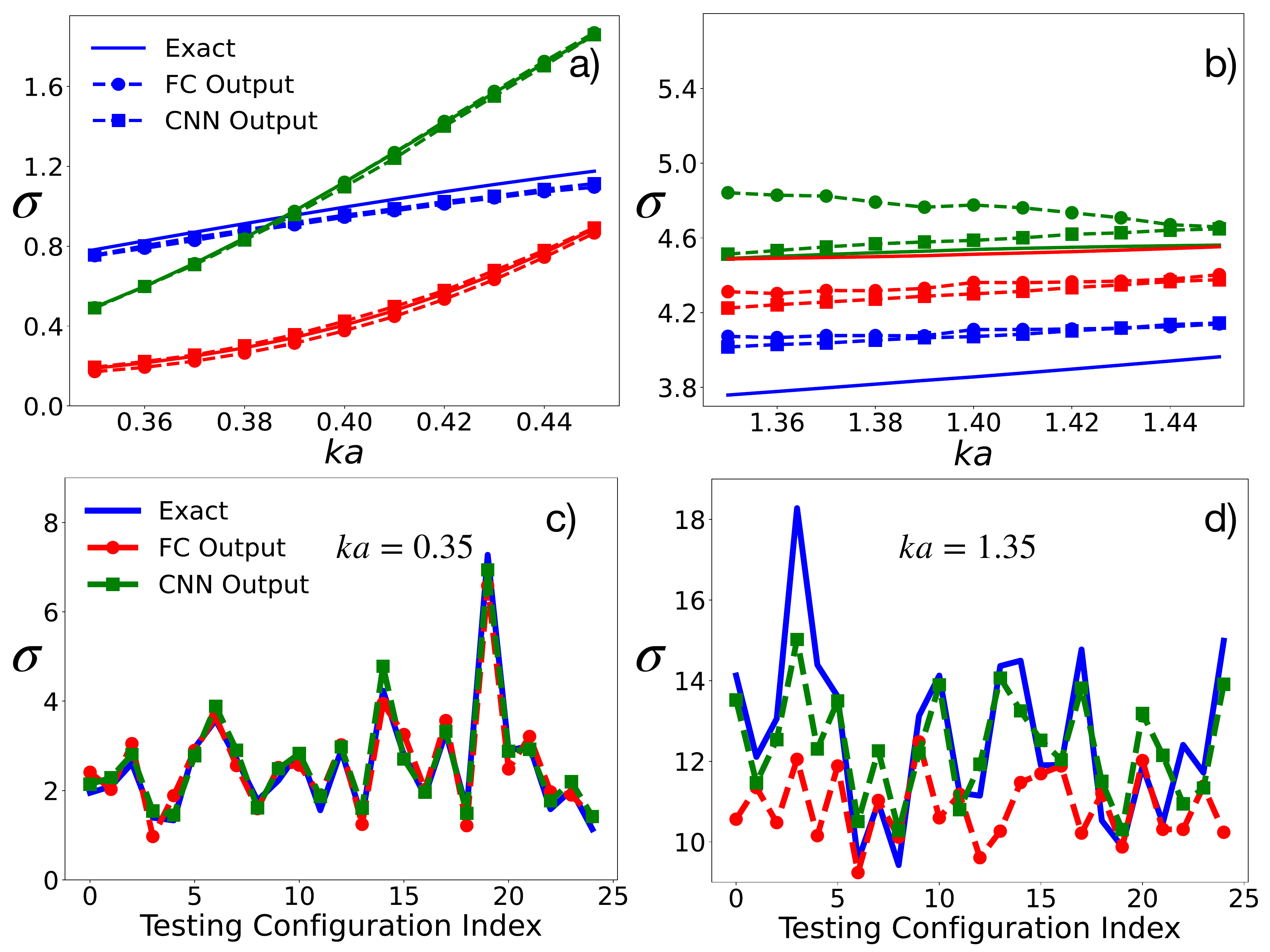}
    \caption{The exact total scattering cross section (TSCS) plotted along with deep learning model 
    predictions. The top row shows the TSCS for three samples with $M=2$, evaluated across $11$ wavenumbers 
    in two regions: (a) $k_i a \in [0.35, 0.45]$ and (b) $k_i a \in [1.35,1.45]$,  where $i=\overline{1,11}$. 
    The bottom row shows the TSCS of $25$ samples with $M=6$, evaluated at single wavenumbers (c) $ka = 0.35$ 
    and (d) $ka = 1.35$. Note that these are separate models trained in two different regions of wavenumbers, 
    and in general the lower $ka$ region tend to produce more accurate approximation.}
    \label{fig4}
\end{figure}

Figure~\ref{fig4} shows the training results with the top row plotting the TSCS when $M=2$ as a function 
of the wavenumber for three random sample images, and the bottom row displaying the output when $M=6$ at 
two different wavenumbers across many samples. We experimented with both FC and CNN in forward design.  
Fig.~\ref{fig4}(a) and~\ref{fig4}(b) show results for each of the two wavenumber regions highlighted in 
Fig.~\ref{fig1}(c). Each color in those panels represents results for a different binary image of the 
scatterer configurations, and different lines distinguish the exact results from those obtained through 
the FC or the CNN. Note that the models are trained separately in the two regions of the wavenumber: 
$k_ia \in [0.35,045]$ and $k_ia \in [1.35, 1.45]$. 

As can be seen, the network performs worse in the larger wavenumber region [e.g., in Fig.~\ref{fig4}(b)]. 
The decrease in the prediction accuracy can be explained by the relatively small variation in TSCS, both 
from a configuration to another and also across $ka$ in that region, making it harder for the network that 
now has to  learn subtle differences. This fact can be inferred from the distributions shown in 
Figs.~\ref{fig1}(e)-(h). In the large wavenumber region [Figs.~\ref{fig1}(f) and~\ref{fig1}(h)], the 
distributions of TSCS in $ka$ are narrower, and also more similar at different $ka$, than in the smaller 
wavenumber region [Figs.~\ref{fig1}(e) and~\ref{fig1}(g)]. We also observe that for these small $M$, FC 
and CNN have more or less the same level of performance.

\begin{figure}[htbp]
    \centering\includegraphics[width=0.6\linewidth]{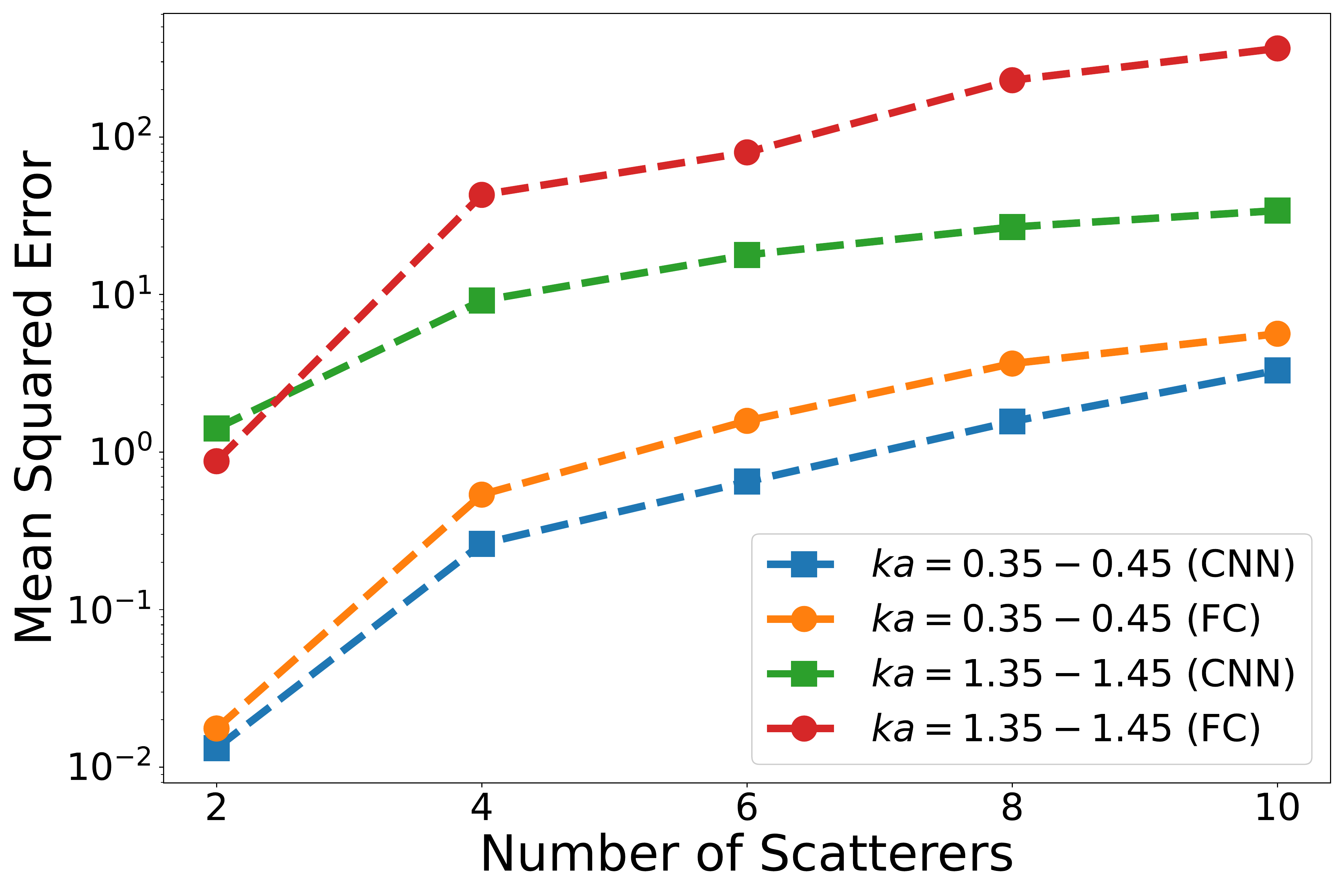}
    \caption{The average mean squared error between exact and predicted TSCS, computed over the testing 
    set of $6,000$ samples in two regions of wavenumber. Note that the error is greater when training 
    with larger wavenumber and with greater numbers of cylinders. This figure also clearly shows that 
    the CNN generally outperforms the FC neural network.}
    \label{fig5}
\end{figure}

Similar trends are seen in Figs.~\ref{fig4}(c) and~\ref{fig4}(d) across 25 different input configurations 
with $M=6$ and at a select $ka$ in each of the wavenumber regions. These plots make it clear that as $M$ 
increases, the CNN  generally outperforms FC, as we expected.

In general, the training accuracy decreases as the number of cylinders increases. This can be seen in 
Fig.~\ref{fig5}, where the average mean squared error, $\sum_{i}^N{(y_{i}^{predicted} - y_{i}^{exact})^2}/N$, 
where $N=6,000$ is the total number of testing samples, is shown. As expected, the difference in error 
between CNN and FC is slightly greater in the larger $ka$ region. In training the neural network, we have 
noticed that the network is prone to overfitting when $M \geq 6$, so in those cases we have added a dropout 
layer, which effectively reduced the numbers of neurons during training. We also observe that CNNs were 
more resistant to overfitting and generally outperformed FCs.

\subsection*{Variational Autoencoders and the Inverse Design}

\begin{figure}[h!]
    \centering\includegraphics[width=0.97\linewidth]{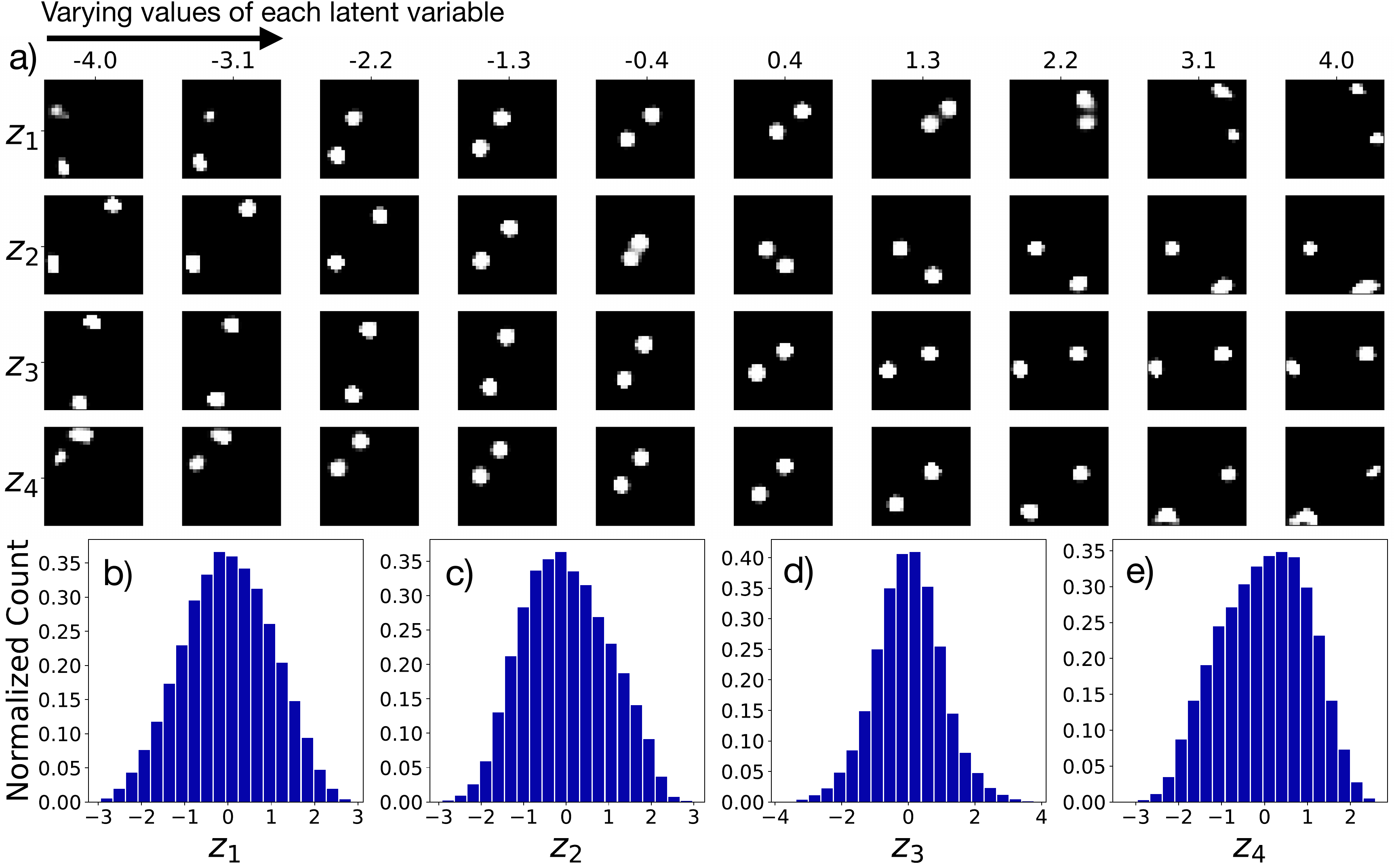}
    \caption{Change of scatterer configurations upon changing each of the latent variables $z_i$ of a 
    trained VAE. We have used the trained decoder to generate new images for the $M=2$ case by sampling 
    from the $z_i$ distributions. The first row of panel (a) shows what happens to the two cylinders 
    when $z_1$ increases while the other three latent variables remain unchanged. The next three rows 
    corresponds to varying $z_2, z_3$ and $ z_4$. The distance between the two cylinders changes as we 
    vary each latent variable, along with the angle of the line connecting them. Panels (b)-(e) show the 
    histograms of each latent variable, which follow the normal distribution as expected.} 
    \label{fig6}
\end{figure}

While it is difficult to interpret what exactly each latent variable represents after training a VAE, 
Fig.~\ref{fig6}(a) attempts to illustrate what attributes of the image the network has learned for the 
case of $M=2$, in which the latent space has only $4$ elements. For example, the first row shows what 
happens to the location of the two cylinders as the value of $z_1$ changes while other latent values 
remain fixed. The second row demonstrates the change of locations as only $z_2$ varies, and so on. 
Even though varying any of the latent values results in a slight difference in location of the scatterers, 
we find that, for example, $z_2$ values are mostly associated with the distance between the two cylinders. 
Increasing negative $z_2$ values decreases this distance. Then, upon the change of sign of $z_2$, the 
orientation of two cylinders changes suddenly before the distance between them increasing slightly again 
with further increasing $z_2$. $z_3$ appears to be responsible for a simultaneous change of distance 
between the cylinders and their rotation while increasing $z_4$ seems to increase that distance and 
uniformly translate the cylinders at the same time.

Figure~\ref{fig6}(b)-(e) are the distributions of the encoded latent variables of $6,000$ testing samples. 
Each latent variable follows roughly a normal distribution with mean and variance close to zero and one 
as expected, since the values of $z$ were sampled from a multivariate normal distribution during training. 
Whereas only $4$ latent variables were needed to achieve a high training accuracy for $M=2$, the length 
of the latent vector must be at least $40$ to reach a reasonable accuracy for $M \geq 8$. In general, 
the more latent variables, the better the accuracy. However, optimization with a large number of variables 
could be time consuming.

\begin{figure}[h!]
    \centering\includegraphics[width=0.97\linewidth]{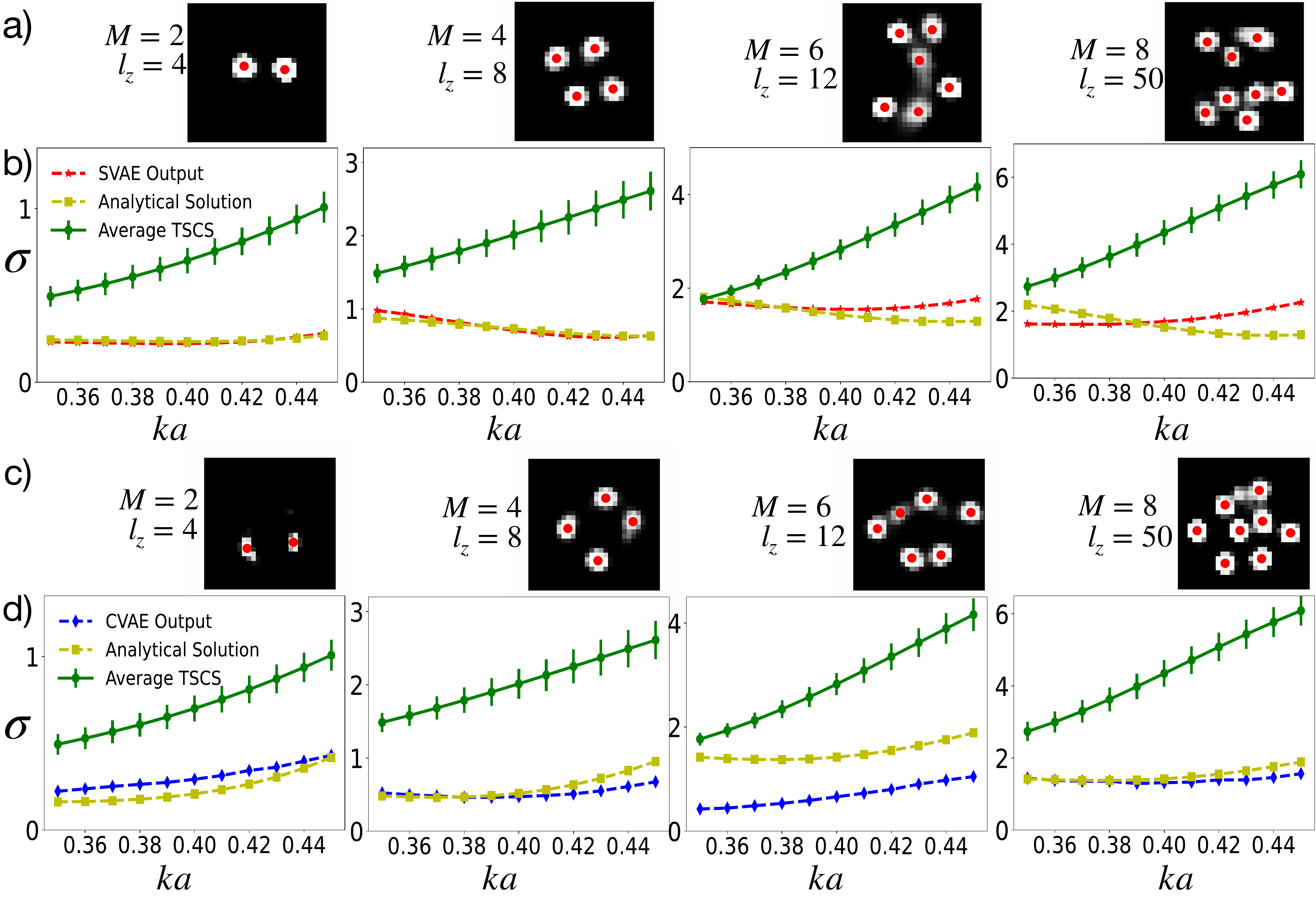}
    \caption{Generated samples from variational autoencoder models and their TSCS for $M=2,4,6$ and 
    $8$. (a) and (c) are binary images produced by optimizing the latent space of each respected 
    configuration of $M$ scatterers using SVAE and CVAE. The number of latent variables used in training the models 
    and optimization is listed on the left side of each image. The red dots, which are obtained by 
    thresholding the decoded images and finding the center of mass among the pixel values, indicate 
    the positions of the cylinders. (b) shows the SVAE output, which is the same as the TSCS regressor 
    output. (d) shows the CVAE output computed by the trained CNN model from forward desired. The 
    average TSCS for each wavenumber $ka$ of $30$ randomly selected training examples and their standard 
    error is also plotted for comparison. While the semi-analytical solution does not match the regressor 
    output completely, the optimized configurations clearly have much lower TSCS compared to the average 
    TSCS.}
    \label{fig7}
\end{figure}

After obtaining the continuous representation of the images through the latent space, we use a Gaussian 
process to search for optimized configurations within the latent space. The optimal set of latent 
variables must reflect both a high probability of being drawn from the normal distribution and a low TSCS.

In designing the architecture for SVAE, we find that the TSCS regressor, when trained to predict the TSCS 
of more than one wavenumber, surprisingly performed better than when trained to predict only the TSCS of 
a single wavenumber. This is a promising sign because it suggests that expanding our region of interest 
in $ka$ [see Fig.~\ref{fig1}(c)] in future studies can potentially improve the performance of our approach.

Figures~\ref{fig7}(a) and \ref{fig7}(b) show the optimization results for $4$ different $M$'s using the SVAE. 
The Gaussian process generated many samples with low TSCS, but here we showcase only one representative sample. 
The numbers of latent variables used for the optimization are $l_z = 4, 8, 12, 50$ for $M = 2, 4, 6, 8$, 
respectively. The optimal latent values are decoded into images shown in Fig.~\ref{fig7}(a) by the SVAE's 
trained decoder. As shown in Fig.~\ref{fig7}(b), while the average TSCS over all the training images (green curve) 
generally increases with $ka$, the SVAE is able to produce new configurations with constant or slightly 
decreasing TSCS, as indicated by the red curve.

To ensure the SVAE did not simply memorize the training data, we searched for similar configurations from 
the $60,000$ samples used to train our models. We compared the generated images and observed that the majority 
of generated configurations, except for the case of $M = 2$, were indeed distinct variations of the existing 
training data. We note that a small change in position of only one cylinder could result in a drastic 
change in the TSCS output.

The locations of the cylinders in generated images, marked by the red dots, are estimated by thresholding 
each image within an appropriate pixel range and calculating the center of mass of the pixel values. The 
centroid positions are then used to compute the analytical solution of the TSCS. As can be seen in 
Fig.~\ref{fig7}(b), the SVAE output is in good agreement with the semi-analytical solution.

Fig.~\ref{fig7}(c) and (d) show optimization results for CVAE. To prompt the decoder to generate images 
with TSCS close to zero, we input $11$ zeros as the conditional vector $c$ of Eq.~\eqref{DKLc}. However, 
we find that the CVAE does not independently produce images with low TSCS this way; the additional 
minimization step through a surrogate model is required. Therefore, similarly to the case of SVAE, we 
use GP to search for possible latent values (excluding the conditional vector) with the lowest TSCS. 
The only difference is that the TSCS regressor in SVAE is replaced by the combination of the decoder 
and the trained CNN from forward design. As can be seen in Fig.~\ref{fig7}(c) and (d), the results are 
encouraging and comparable to those obtained through the SVAE.

We observe that generally, optimizing the objective function with larger numbers of latent variables 
can result in clearer and more realistic samples but also takes a longer time. This can be inferred from 
sample images in Fig.~\ref{fig7}(a) and (c), where those corresponding to $M = 8$ seem to be sharper 
than the images for $M = 6$ despite having more cylinders. This is at least partly due to the fact that 
the VAE models for $M = 8$ were trained with greater number of latent variables ($l_z = 50$) compared to 
$l_z = 12$ for $M = 6$. For $l_z \leq 50$, the objective function converges to a minimum value within a 
few minutes using the Gaussian process optimization.

In analyzing the resolution of the generated images, we noticed that some of the cylinders, represented 
by the white regions, are blurry or have lower pixel values, suggesting that these cylinders might not 
be as important as others. This is interesting since it shows that the machines might be able to predict 
different size cylinders for optimum results. We also computed the TSCS of these images with the blurry 
discs removed, and found that sometimes the semi-analytical solutions were more consistent with the CNN and 
the regressor output. This proves that the VAEs could even generate optimal configurations with different 
numbers of cylinders, even though the models are trained with data of only one specific $M$ value.

\section*{Conclusions}

We have demonstrated a new method for inverse design of planar configurations of rigid cylinders with 
specific scattering properties at different wavenumbers. Our method, which is based on a combination of 
VAE and supervised learning approaches, indicate that it is possible to eliminate the traditional, time 
consuming, gradient based optimizations in acoustic metamaterial design. Our goal here has been to design 
scatterer configurations that produce minimal TSCS.

The continuous encoded representation of the scatterers via VAEs permits Gaussian process optimization 
within a few minutes to search for configurations with the lowest TSCS. Many of such generated samples 
have low, and sometimes decreasing, TSCS, which is seen as a different trend when compared to the average 
of existing training data. We observed that, at least in the case of two scatterers, our SVAE model was 
able to learn the TSCS dependency on the relative position vectors ${\bf r}_{ij}={\bf r}_{i}-{\bf r}_{j}$ 
\cite{Amirkulova2020}.

We find that the training accuracy of the  CNN regressor decreases with increasing the number of scatterers 
and wavenumbers. To further improve the training accuracy, the training data needs to be carefully selected 
to maximize the variance of the TSCS across all wavenumbers. Considering nonuniform sets of cylinders with 
different radii can provide another possible way to decrease TSCS. Working with such nonuniform cylinders 
may also produce higher variance in the TSCS.

We also find that the TSCS regressor in our SVAE performs better when trained at more than one value of 
$ka$, a behavior that is the opposite of what we observe for the CNN model in our forward design. It is 
unclear why there is such peculiarity in the predictive power between the forward and inverse designs. 
Nevertheless, the difference suggests that training the model with more $ka$ values might result in a 
more robust and comprehensive generative model, which should be further explored in future studies.

\section*{List of abbreviations}
\noindent TSCS -  total scattering cross section;\\
FC - fully connected;\\
CNN - convolutional neural network;\\
ReLU - rectified linear unit;\\
KL - Kullback-Leibler;\\
VAE - variational autoencoder;\\
CVAE - conditional variational autoencoder;\\
SVAE - supervised variational autoencoder;

\section*{Availability of data and materials}
The datasets used and/or analysed during the current study are available from the corresponding author 
on reasonable request.

\section*{Competing interests}
The authors declare that they have no competing interests

\section*{Funding}
The proposed model was developed under 
Small Grant Project (SGP) grant from San Jose State University. The SGP grant supported TT, FA, and EK in 
the design of the study, model training, data collection, analysis, interpretation of data and in writing the manuscript.

\section*{Authors' contributions}
Conceptualization, F.A. and E.K.; methodology, F.A. and E.K.; software, T.T., F.A. and E.K.; validation, 
T.T., F.A. and E.K.; formal analysis, T.T., F.A. and E.K.; investigation, T.T., F.A. and E.K.; resources 
F.A. and E.K.; data curation, F.A.;  writing---original draft preparation, T.T., F.A. and E.K.; visualization, 
T.T. ; supervision, F.A. and E.K.; project administration, F.A.; funding acquisition, F.A. and E.K. 
All authors have read and agreed to the published version of the manuscript.

\section*{Acknowledgements
}
TT, FA, and EK acknowledge Small Grant Project (SGP) grant support from San Jose State University.

\section*{Appendix}

\begin{figure}[h!]
    \centering\includegraphics[width=0.9\linewidth]{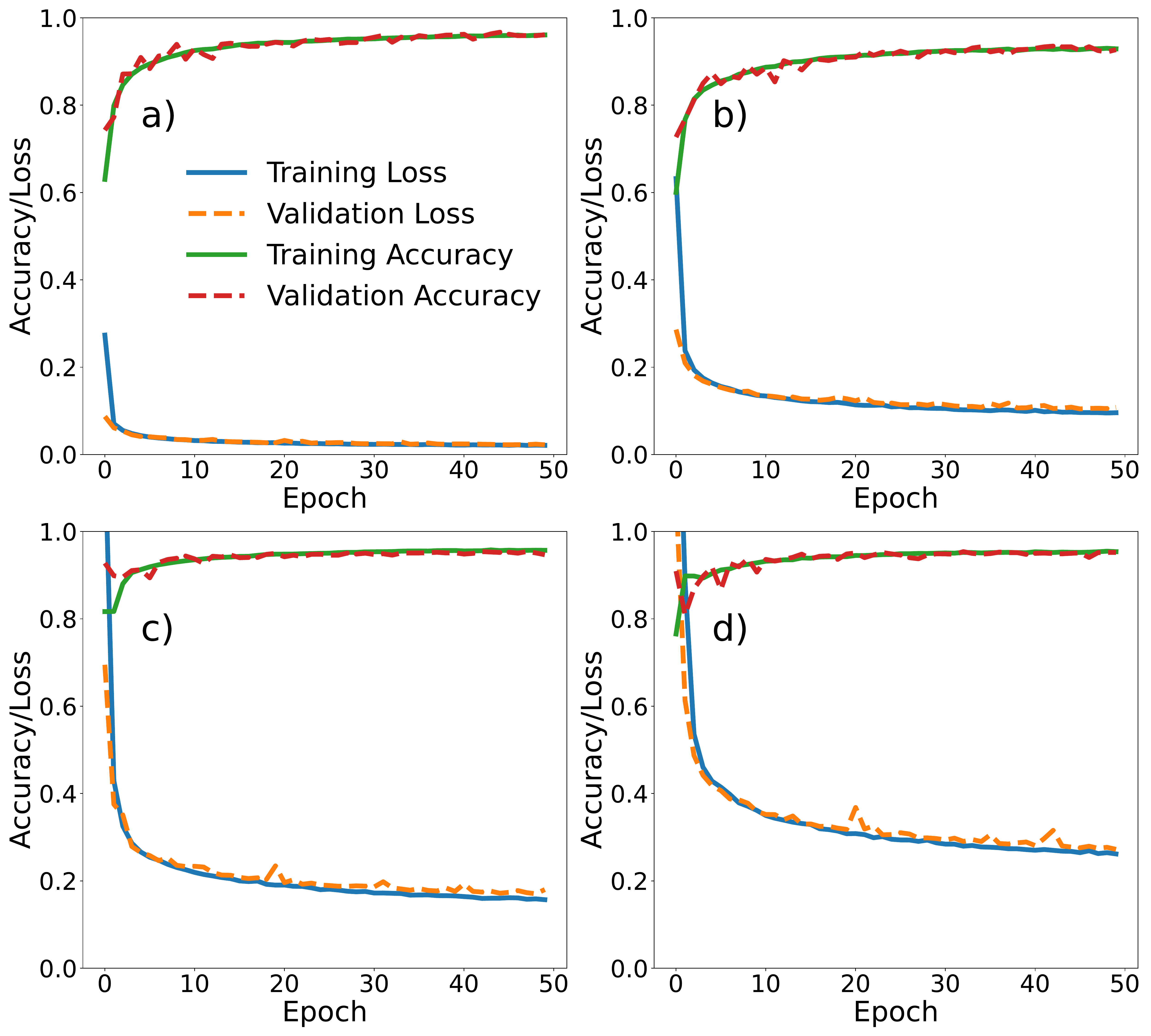}
    \caption{The CNN training progression for (a) $M=2$ (b) $M=4$ (c) $M=6$ (d) $M=8$ in region 
    $k_i a \in [0.35, 0.45]$. We used the Adam optimizer with a $0.001$ learning rate and mean squared error 
    as the loss function. Each epoch was trained on a small batch of the data with size of either $128$ or 
    $256$ samples, depending on the numbers of cylinders, larger batch size and smaller learning rate is 
    necessary to prevent over fitting. The top two curves show the training and validation accuracy, along 
    with the losses at the bottom two curves.}
    \label{fig8}
\end{figure}

Figure~\ref{fig8} shows the training progression of CNN in the forward design. The training and validation 
loss converge after about 40 epochs. An epoch is the time it takes for the network to go over the entire 
training data, $10\%$ of which was set aside to validate the network's performance on unseen data during 
training. The loss at each epoch is greater when training with larger $M$ values. Further optimization of 
the network architecture and hyperparameters could improve training accuracy for large $M$.

\begin{figure}[h!]
    \centering\includegraphics[width=0.9\linewidth]{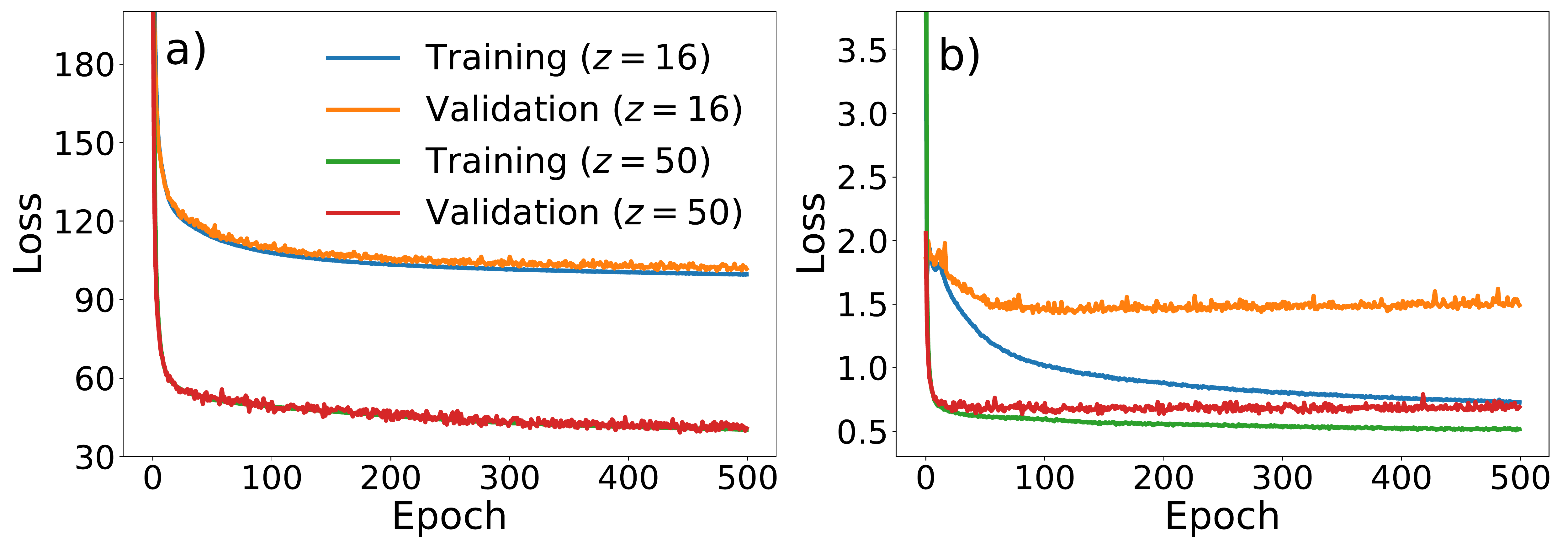}
    \caption{SAVE losses for (a) the reconstruction task and (b) the TSCS regression task for training in 
    the region $k a \in [0.35, 0.45]$ when $M=8$. The TSCS regression loss tends to converge faster than 
    the reconstruction loss, however, it shows signs of overfitting. Both losses are much lower when training 
    with a larger $l_z$.}
    \label{fig9}
\end{figure}

Figure~\ref{fig9} shows training progression of SVAE for $M = 8$ using two different lengths of the latent 
vector $z = 16$ and $50$. The losses for both reconstruction task and TSCS regression task are much lower 
for $z = 50$. Note that the regression loss for $z = 16$ in Fig.~\ref{fig9}(b) exhibits signs of overfitting 
since the validation loss stops decreasing with the training loss after epoch $20$ to $30$. Meanwhile, 
the losses for the reconstruction task are still declining. The overfitting affect of the TSCS regressor 
could be reduced by training with more latent variables, a smaller learning rate or a larger drop out rate.


\begin{backmatter}

\bibliographystyle{bmc-mathphys}


\newcommand{\BMCxmlcomment}[1]{}

\BMCxmlcomment{

<refgrp>

<bibl id="B1">
  <title><p>{Metamaterials beyond electromagnetism}</p></title>
  <aug>
    <au><snm>Kadic</snm><fnm>M</fnm></au>
    <au><snm>B\"{u}ckmann</snm><fnm>T</fnm></au>
    <au><snm>Schittny</snm><fnm>R</fnm></au>
    <au><snm>Wegener</snm><fnm>M</fnm></au>
  </aug>
  <source>Rep. Prog. Phys.</source>
  <pubdate>2013</pubdate>
  <volume>76</volume>
  <issue>12</issue>
  <fpage>126501+</fpage>
</bibl>

<bibl id="B2">
  <title><p>{Comparative study of potential pentamodal metamaterials inspired
  by Bravais lattices}</p></title>
  <aug>
    <au><snm>M\'{e}jica</snm><fnm>GF</fnm></au>
    <au><snm>Lantada</snm><fnm>AD</fnm></au>
  </aug>
  <source>Smart Mat. Struct.</source>
  <pubdate>2013</pubdate>
  <volume>22</volume>
  <issue>11</issue>
  <fpage>115013+</fpage>
</bibl>

<bibl id="B3">
  <title><p>Focusing, refraction, and asymmetric transmission of elastic waves
  in solid metamaterials with aligned parallel gaps</p></title>
  <aug>
    <au><snm>Su</snm><fnm>X</fnm></au>
    <au><snm>Norris</snm><fnm>AN</fnm></au>
  </aug>
  <source>The Journal of the Acoustical Society of America</source>
  <publisher>Acoustical Society of America (ASA)</publisher>
  <pubdate>2016</pubdate>
  <volume>139</volume>
  <issue>6</issue>
  <fpage>3386</fpage>
  <lpage>-3394</lpage>
  <url>http://dx.doi.org/10.1121/1.4950770</url>
</bibl>

<bibl id="B4">
  <title><p>Acoustic Poisson-like effect in periodic structures</p></title>
  <aug>
    <au><snm>Titovich</snm><fnm>AS</fnm></au>
    <au><snm>Norris</snm><fnm>AN</fnm></au>
  </aug>
  <source>The Journal of the Acoustical Society of America</source>
  <publisher>Acoustical Society of America ({ASA})</publisher>
  <pubdate>2016</pubdate>
  <volume>139</volume>
  <issue>6</issue>
  <fpage>3353</fpage>
  <lpage>-3356</lpage>
</bibl>

<bibl id="B5">
  <title><p>Metaclusters for the Full Control of Mechanical Waves</p></title>
  <aug>
    <au><snm>Packo</snm><fnm>P</fnm></au>
    <au><snm>Norris</snm><fnm>AN</fnm></au>
    <au><snm>Torrent</snm><fnm>D</fnm></au>
  </aug>
  <source>Physical Review Applied</source>
  <publisher>American Physical Society ({APS})</publisher>
  <pubdate>2021</pubdate>
  <volume>15</volume>
  <issue>1</issue>
</bibl>

<bibl id="B6">
  <title><p>Acoustic cloaking theory</p></title>
  <aug>
    <au><snm>Norris</snm><fnm>A. N.</fnm></au>
  </aug>
  <source>Proc. R. Soc. A</source>
  <pubdate>2008</pubdate>
  <volume>464</volume>
  <fpage>2411</fpage>
  <lpage>-2434</lpage>
</bibl>

<bibl id="B7">
  <title><p>The gradient of total multiple scattering cross-section and its
  application to acoustic cloaking</p></title>
  <aug>
    <au><snm>Amirkulova</snm><fnm>FA</fnm></au>
    <au><snm>Norris</snm><fnm>AN</fnm></au>
  </aug>
  <source>Journal of Theoretical and Computational Acoustics</source>
  <publisher>World Scientific Pub Co Pte Lt</publisher>
  <pubdate>2020</pubdate>
  <fpage>1950016</fpage>
</bibl>

<bibl id="B8">
  <title><p>A high transmission broadband gradient index lens using elastic
  shell acoustic metamaterial elements</p></title>
  <aug>
    <au><snm>Titovich</snm><fnm>A. S.</fnm></au>
    <au><snm>Norris</snm><fnm>A.N.</fnm></au>
    <au><snm>Haberman</snm><fnm>M. R.</fnm></au>
  </aug>
  <source>Journal of the Acoustical Society of America</source>
  <pubdate>2016</pubdate>
  <volume>139</volume>
  <fpage>3357</fpage>
  <lpage>-3364</lpage>
</bibl>

<bibl id="B9">
  <title><p>Broadband acoustic metamaterial design using gradient-based
  optimization</p></title>
  <aug>
    <au><snm>Fahey</snm><fnm>L</fnm></au>
    <au><snm>Amirkulova</snm><fnm>F</fnm></au>
    <au><snm>Norris</snm><fnm>A</fnm></au>
  </aug>
  <source>The Journal of the Acoustical Society of America</source>
  <publisher>Acoustical Society of America ({ASA})</publisher>
  <pubdate>2019</pubdate>
  <volume>146</volume>
  <issue>4</issue>
  <fpage>2830</fpage>
  <lpage>-2830</lpage>
</bibl>

<bibl id="B10">
  <title><p>Acoustic Cloaking with Plasmonic Shells</p></title>
  <aug>
    <au><snm>Haberman</snm><fnm>MR</fnm></au>
    <au><snm>Guild</snm><fnm>MD</fnm></au>
    <au><snm>Al\`u</snm><fnm>A</fnm></au>
  </aug>
  <source>Acoustic Metamaterials</source>
  <publisher>New York London: Springer</publisher>
  <editor>Richard V. Craster and Sebastien Guenneau</editor>
  <series><title><p>Springer Series in Materials Science</p></title></series>
  <pubdate>2013</pubdate>
  <volume>166</volume>
  <fpage>241</fpage>
  <lpage>265</lpage>
</bibl>

<bibl id="B11">
  <title><p>Topology Optimized Cloak for Airborne Sound</p></title>
  <aug>
    <au><snm>Andkj{\ae}r</snm><fnm>J.</fnm></au>
    <au><snm>Sigmund</snm><fnm>O.</fnm></au>
  </aug>
  <source>Journal of Vibration and Acoustics</source>
  <pubdate>2013</pubdate>
  <volume>135</volume>
  <fpage>041011</fpage>
</bibl>

<bibl id="B12">
  <title><p>Directional acoustic source by scattering acoustical
  elements</p></title>
  <aug>
    <au><snm>H\r{a}kansson</snm><fnm>A</fnm></au>
    <au><snm>Torrent</snm><fnm>D</fnm></au>
    <au><snm>Cervera</snm><fnm>F</fnm></au>
    <au><snm>S\'anchez Dehesa</snm><fnm>J</fnm></au>
  </aug>
  <source>Appl. Phys. Lett.</source>
  <publisher>AIP Publishing</publisher>
  <pubdate>2007</pubdate>
  <volume>90</volume>
  <issue>22</issue>
  <fpage>224107</fpage>
  <url>http://dx.doi.org/10.1063/1.2743947</url>
</bibl>

<bibl id="B13">
  <title><p>Noise control by sonic crystal barriers made of recycled
  materials</p></title>
  <aug>
    <au><snm>Saanchez Dehesa</snm><fnm>J</fnm></au>
    <au><snm>Garcia Chocano</snm><fnm>VM</fnm></au>
    <au><snm>Torrent</snm><fnm>D</fnm></au>
    <au><snm>Cervera</snm><fnm>F</fnm></au>
    <au><snm>Cabrera</snm><fnm>S</fnm></au>
    <au><snm>Simon</snm><fnm>F</fnm></au>
  </aug>
  <source>J. Acoust. Soc. Am.</source>
  <publisher>Acoustical Society of America (ASA)</publisher>
  <pubdate>2011</pubdate>
  <volume>129</volume>
  <issue>3</issue>
  <fpage>1173</fpage>
  <url>http://dx.doi.org/10.1121/1.3531815</url>
</bibl>

<bibl id="B14">
  <title><p>Design of multi-directional acoustic cloaks using two-dimensional
  shape optimization and the boundary element method</p></title>
  <aug>
    <au><snm>Andersen</snm><fnm>P.</fnm></au>
    <au><snm>Henriquez</snm><fnm>V.</fnm></au>
    <au><snm>Sanchis</snm><fnm>L.</fnm></au>
    <au><snm>Sánchez Dehesa</snm><fnm>J.</fnm></au>
  </aug>
  <source>Proc. of ICA 2019 AND EAA EUROREGIO Deutsche Gesellschaft für
  Akustik e.V.</source>
  <pubdate>2019</pubdate>
  <fpage>5600</fpage>
  <lpage>5606</lpage>
</bibl>

<bibl id="B15">
  <title><p>Review of numerical optimization techniques for meta-device design
  [Invited]</p></title>
  <aug>
    <au><snm>Campbell</snm><fnm>S.</fnm></au>
    <au><snm>Sell</snm><fnm>D.</fnm></au>
    <au><snm>Jenkins R.</snm><fnm>E</fnm></au>
    <au><snm>Fan</snm><fnm>J</fnm></au>
    <au><snm>Werner</snm><fnm>D.</fnm></au>
  </aug>
  <source>Optical Materials Express</source>
  <pubdate>2019</pubdate>
  <volume>9</volume>
  <issue>4</issue>
  <fpage>1842</fpage>
  <lpage>1863</lpage>
</bibl>

<bibl id="B16">
  <title><p>Numerical Optimization Methods for Metasurfaces</p></title>
  <aug>
    <au><snm>Elsawy</snm><fnm>MMR</fnm></au>
    <au><snm>Lanteri</snm><fnm>S</fnm></au>
    <au><snm>Duvigneau</snm><fnm>R</fnm></au>
    <au><snm>Fan</snm><fnm>JA</fnm></au>
    <au><snm>Genevet</snm><fnm>P</fnm></au>
  </aug>
  <source>Laser {\&} Photonics Reviews</source>
  <publisher>Wiley</publisher>
  <pubdate>2020</pubdate>
  <volume>14</volume>
  <issue>10</issue>
  <fpage>1900445</fpage>
</bibl>

<bibl id="B17">
  <title><p>A {GPU}-Accelerated Machine Learning Framework for Molecular
  Simulation: Hoomd-Blue with {TensorFlow}</p></title>
  <aug>
    <au><snm>Barrett</snm><fnm>R</fnm></au>
    <au><snm>Chakraborty</snm><fnm>M</fnm></au>
    <au><snm>Amirkulova</snm><fnm>D</fnm></au>
    <au><snm>Gandhi</snm><fnm>H</fnm></au>
    <au><snm>White</snm><fnm>A</fnm></au>
  </aug>
  <source>10.26434/chemrxiv.8019527</source>
  <publisher>American Chemical Society ({ACS})</publisher>
  <pubdate>2019</pubdate>
</bibl>

<bibl id="B18">
  <title><p>Deep learning for molecular design{\textemdash}a review of the
  state of the art</p></title>
  <aug>
    <au><snm>Elton</snm><fnm>DC</fnm></au>
    <au><snm>Boukouvalas</snm><fnm>Z</fnm></au>
    <au><snm>Fuge</snm><fnm>MD</fnm></au>
    <au><snm>Chung</snm><fnm>PW</fnm></au>
  </aug>
  <source>Molecular Systems Design {\&} Engineering</source>
  <publisher>Royal Society of Chemistry ({RSC})</publisher>
  <pubdate>2019</pubdate>
  <volume>4</volume>
  <issue>4</issue>
  <fpage>828</fpage>
  <lpage>-849</lpage>
</bibl>

<bibl id="B19">
  <title><p>Deep Neural Network Inverse Design of Integrated Photonic Power
  Splitters</p></title>
  <aug>
    <au><snm>Tahersima</snm><fnm>MH</fnm></au>
    <au><snm>Kojima</snm><fnm>K</fnm></au>
    <au><snm>Koike Akino</snm><fnm>T</fnm></au>
    <au><snm>Jha</snm><fnm>D</fnm></au>
    <au><snm>Wang</snm><fnm>B</fnm></au>
    <au><snm>Lin</snm><fnm>C</fnm></au>
    <au><snm>Parsons</snm><fnm>K</fnm></au>
  </aug>
  <source>Scientific Reports</source>
  <publisher>Springer Nature</publisher>
  <pubdate>2019</pubdate>
  <volume>9</volume>
  <issue>1</issue>
  <fpage>1368</fpage>
</bibl>

<bibl id="B20">
  <title><p>Deep learning enabled inverse design in nanophotonics</p></title>
  <aug>
    <au><snm>So</snm><fnm>S</fnm></au>
    <au><snm>Badloe</snm><fnm>T</fnm></au>
    <au><snm>Noh</snm><fnm>J</fnm></au>
    <au><snm>Rho</snm><fnm>J</fnm></au>
    <au><snm>Bravo Abad</snm><fnm>J</fnm></au>
  </aug>
  <source>Nanophotonics</source>
  <publisher>Walter de Gruyter {GmbH}</publisher>
  <pubdate>2020</pubdate>
  <volume>9</volume>
  <issue>5</issue>
  <fpage>1041</fpage>
  <lpage>-1057</lpage>
</bibl>

<bibl id="B21">
  <title><p>Generative adversarial networks for the design of acoustic
  metamaterials</p></title>
  <aug>
    <au><snm>Gurbuz</snm><fnm>C</fnm></au>
    <au><snm>Kronowetter</snm><fnm>F</fnm></au>
    <au><snm>Dietz</snm><fnm>C</fnm></au>
    <au><snm>Eser</snm><fnm>M</fnm></au>
    <au><snm>Schmid</snm><fnm>J</fnm></au>
    <au><snm>Marburg</snm><fnm>S</fnm></au>
  </aug>
  <source>The Journal of the Acoustical Society of America</source>
  <publisher>Acoustical Society of America ({ASA})</publisher>
  <pubdate>2021</pubdate>
  <volume>149</volume>
  <issue>2</issue>
  <fpage>1162</fpage>
  <lpage>-1174</lpage>
</bibl>

<bibl id="B22">
  <title><p>Reinforcement learning applied to metamaterial design</p></title>
  <aug>
    <au><snm>Shah</snm><fnm>T</fnm></au>
    <au><snm>Zhuo</snm><fnm>L</fnm></au>
    <au><snm>Lai</snm><fnm>P</fnm></au>
    <au><snm>Rosa Moreno</snm><fnm>ADL</fnm></au>
    <au><snm>Amirkulova</snm><fnm>F</fnm></au>
    <au><snm>Gerstoft</snm><fnm>P</fnm></au>
  </aug>
  <source>Journal of the Acoustical Society of America</source>
  <pubdate>2021 https://doi.org/10.1121/10.0005545</pubdate>
  <volume>150</volume>
  <issue>1</issue>
</bibl>

<bibl id="B23">
  <title><p>Design of one-dimensional acoustic metamaterials using machine
  learning and cell concatenation</p></title>
  <aug>
    <au><snm>Wu</snm><fnm>RT</fnm></au>
    <au><snm>Liu</snm><fnm>TW</fnm></au>
    <au><snm>Jahanshahi</snm><fnm>MR</fnm></au>
    <au><snm>Semperlotti</snm><fnm>F</fnm></au>
  </aug>
  <source>Structural and Multidisciplinary Optimization</source>
  <publisher>Springer Science and Business Media {LLC}</publisher>
  <pubdate>2021</pubdate>
  <volume>63</volume>
  <issue>5</issue>
  <fpage>2399</fpage>
  <lpage>-2423</lpage>
</bibl>

<bibl id="B24">
  <title><p>A physics-constrained deep learning based approach for acoustic
  inverse scattering problems</p></title>
  <aug>
    <au><snm>Wu</snm><fnm>RT</fnm></au>
    <au><snm>Jokar</snm><fnm>M</fnm></au>
    <au><snm>Jahanshahi</snm><fnm>MR</fnm></au>
    <au><snm>Semperlotti</snm><fnm>F</fnm></au>
  </aug>
  <source>Mechanical Systems and Signal Processing</source>
  <publisher>Elsevier {BV}</publisher>
  <pubdate>2022</pubdate>
  <volume>164</volume>
  <fpage>108190</fpage>
</bibl>

<bibl id="B25">
  <title><p>Deterministic and probabilistic deep learning models for inverse
  design of broadband acoustic cloak</p></title>
  <aug>
    <au><snm>Ahmed</snm><fnm>WW</fnm></au>
    <au><snm>Farhat</snm><fnm>M</fnm></au>
    <au><snm>Zhang</snm><fnm>X</fnm></au>
    <au><snm>Wu</snm><fnm>Y</fnm></au>
  </aug>
  <source>Physical Review Research</source>
  <publisher>American Physical Society ({APS})</publisher>
  <pubdate>2021</pubdate>
  <volume>3</volume>
  <issue>1</issue>
</bibl>

<bibl id="B26">
  <title><p>A spherical basis function neural network for approximating
  acoustic scatter</p></title>
  <aug>
    <au><snm>Jenison</snm><fnm>R.L.</fnm></au>
  </aug>
  <source>Journal of the Acoustical Society of America</source>
  <pubdate>1996</pubdate>
  <volume>99</volume>
  <issue>5</issue>
</bibl>

<bibl id="B27">
  <title><p>Models of direction estimation with spherical-function approximated
  cortical receptive fields</p></title>
  <aug>
    <au><snm>Jenison</snm><fnm>R. L.</fnm></au>
  </aug>
  <source>Central Auditory Processing and Neural Modeling</source>
  <publisher>New York: Plenum Press</publisher>
  <editor>Poon, P. and Brugge, J.</editor>
  <pubdate>1998</pubdate>
  <fpage>161</fpage>
  <lpage>17</lpage>
</bibl>

<bibl id="B28">
  <title><p>Neural network model for solving integral equation of acoustic
  scattering using wavelet basis</p></title>
  <aug>
    <au><snm>Hesham</snm><fnm>M.</fnm></au>
    <au><snm>El Gamal</snm><fnm>M.</fnm></au>
  </aug>
  <source>Commun. Numer. Meth. Engng</source>
  <pubdate>2008</pubdate>
  <volume>24</volume>
  <fpage>183–194</fpage>
</bibl>

<bibl id="B29">
  <title><p>Neural-Network Solution of the Nonuniqueness Problem in Acoustic
  Scattering Using Wavelets,</p></title>
  <aug>
    <au><snm>Hesham</snm><fnm>M.</fnm></au>
    <au><snm>El Gamal</snm><fnm>M.</fnm></au>
  </aug>
  <source>Int. J. for Comp.l Methods in Engng Science and Mech.</source>
  <pubdate>2008</pubdate>
  <volume>9</volume>
  <issue>4</issue>
  <fpage>217</fpage>
  <lpage>222</lpage>
</bibl>

<bibl id="B30">
  <title><p>Principal Component Analysis Applied to Gradient Fields in Band Gap
  Optimization Problems for Metamaterials</p></title>
  <aug>
    <au><snm>Gnecco</snm><fnm>G</fnm></au>
    <au><snm>Bacigalupo</snm><fnm>A</fnm></au>
    <au><snm>Fantoni</snm><fnm>F</fnm></au>
    <au><snm>Selvi</snm><fnm>D</fnm></au>
  </aug>
  <source>arXiv:2104.02588 [cs.CE]</source>
  <pubdate>2021</pubdate>
</bibl>

<bibl id="B31">
  <title><p>Fast Acoustic Scattering Using Convolutional Neural
  Networks</p></title>
  <aug>
    <au><snm>Fan</snm><fnm>Z</fnm></au>
    <au><snm>Vineet</snm><fnm>V</fnm></au>
    <au><snm>Gamper</snm><fnm>H</fnm></au>
    <au><snm>Raghuvanshi</snm><fnm>N</fnm></au>
  </aug>
  <source>ICASSP 2020 - 2020 IEEE International Conference on Acoustics, Speech
  and Signal Processing (ICASSP)</source>
  <pubdate>2020</pubdate>
  <fpage>171</fpage>
  <lpage>175</lpage>
</bibl>

<bibl id="B32">
  <title><p>Prediction of Object Geometry from Acoustic Scattering Using
  Convolutional Neural Networks</p></title>
  <aug>
    <au><snm>Fan</snm><fnm>Z</fnm></au>
    <au><snm>Vineet</snm><fnm>V</fnm></au>
    <au><snm>Lu</snm><fnm>C</fnm></au>
    <au><snm>Wu</snm><fnm>T. W.</fnm></au>
    <au><snm>McMullen</snm><fnm>K</fnm></au>
  </aug>
  <source>arXiv:2010.10691</source>
  <pubdate>2021</pubdate>
</bibl>

<bibl id="B33">
  <title><p>Solving a kind of inverse scattering problem of acoustic waves
  based on linear sampling method and neural network</p></title>
  <aug>
    <au><snm>Meng</snm><fnm>P</fnm></au>
    <au><snm>Su</snm><fnm>L</fnm></au>
    <au><snm>Yin</snm><fnm>W</fnm></au>
    <au><snm>Zhang</snm><fnm>S</fnm></au>
  </aug>
  <source>Alexandria Engineering Journal</source>
  <publisher>Elsevier {BV}</publisher>
  <pubdate>2020</pubdate>
  <volume>59</volume>
  <issue>3</issue>
  <fpage>1451</fpage>
  <lpage>-1462</lpage>
</bibl>

<bibl id="B34">
  <title><p>Generative Adversarial Nets</p></title>
  <aug>
    <au><snm>Goodfellow</snm><fnm>IJ</fnm></au>
    <au><snm>Pouget Abadie</snm><fnm>J</fnm></au>
    <au><snm>Mirza</snm><fnm>M</fnm></au>
    <au><snm>Xu</snm><fnm>B</fnm></au>
    <au><snm>Warde Farley</snm><fnm>D</fnm></au>
    <au><snm>Ozair</snm><fnm>S</fnm></au>
    <au><snm>Courville</snm><fnm>A</fnm></au>
    <au><snm>Bengio</snm><fnm>Y</fnm></au>
  </aug>
  <source>Advances in Neural Information Processing Systems 27 (NIPS
  2014)</source>
  <pubdate>2014</pubdate>
</bibl>

<bibl id="B35">
  <title><p>Semi-Supervised Learning with Deep Generative Models</p></title>
  <aug>
    <au><snm>Kingma</snm><fnm>DP</fnm></au>
    <au><snm>Rezende</snm><fnm>DJ</fnm></au>
    <au><snm>Mohamed</snm><fnm>S</fnm></au>
    <au><snm>Welling</snm><fnm>M</fnm></au>
  </aug>
  <source>https://arxiv.org/abs/1406.5298</source>
  <pubdate>2014</pubdate>
</bibl>

<bibl id="B36">
  <title><p>Probabilistic Representation and Inverse Design of Metamaterials
  Based on a Deep Generative Model with Semi-Supervised Learning
  Strategy</p></title>
  <aug>
    <au><snm>Ma</snm><fnm>W</fnm></au>
    <au><snm>Cheng</snm><fnm>F</fnm></au>
    <au><snm>Xu</snm><fnm>Y</fnm></au>
    <au><snm>Wen</snm><fnm>Q</fnm></au>
    <au><snm>Liu</snm><fnm>Y</fnm></au>
  </aug>
  <source>Advanced Materials</source>
  <publisher>Wiley</publisher>
  <pubdate>2019</pubdate>
  <volume>31</volume>
  <issue>35</issue>
  <fpage>1901111</fpage>
</bibl>

<bibl id="B37">
  <title><p>Multiple Scattering: Interaction of Time-harmonic Waves with {N}
  Obstacles</p></title>
  <aug>
    <au><snm>Martin</snm><fnm>P. A.</fnm></au>
  </aug>
  <publisher>New York: Cambridge University Press</publisher>
  <pubdate>2006</pubdate>
</bibl>

<bibl id="B38">
  <title><p>Acoustic integrated extinction</p></title>
  <aug>
    <au><snm>Norris</snm><fnm>A. N.</fnm></au>
  </aug>
  <source>Proc. R. Soc. A</source>
  <pubdate>2015</pubdate>
  <volume>471</volume>
  <issue>2177</issue>
  <fpage>20150008+</fpage>
</bibl>

<bibl id="B39">
  <title><p>Acoustic and Elastic Multiple Scattering and Radiation from
  Cylindrical Structures</p></title>
  <aug>
    <au><snm>Amirkulova</snm><fnm>F. A.</fnm></au>
  </aug>
  <source>PhD thesis</source>
  <publisher>Rutgers University</publisher>
  <pubdate>2014</pubdate>
</bibl>

<bibl id="B40">
  <title><p>Auto-Encoding Variational Bayes</p></title>
  <aug>
    <au><snm>Kingma</snm><fnm>M.</fnm></au>
  </aug>
  <source>arXiv:1312.6114v10</source>
  <pubdate>2014</pubdate>
</bibl>

<bibl id="B41">
  <title><p>Semi-Supervised Learning with Deep Generative Models</p></title>
  <aug>
    <au><snm>Kingma</snm><fnm>DP</fnm></au>
    <au><snm>Rezende</snm><fnm>DJ</fnm></au>
    <au><snm>Mohamed</snm><fnm>S</fnm></au>
    <au><snm>Welling</snm><fnm>M</fnm></au>
  </aug>
  <pubdate>2014</pubdate>
</bibl>

<bibl id="B42">
  <title><p>Learning structured output representation using deep conditional
  generative models</p></title>
  <aug>
    <au><snm>Sohn</snm><fnm>K</fnm></au>
    <au><snm>Lee</snm><fnm>H</fnm></au>
    <au><snm>Yan</snm><fnm>X</fnm></au>
  </aug>
  <source>Advances in neural information processing systems</source>
  <pubdate>2015</pubdate>
  <volume>28</volume>
  <fpage>3483</fpage>
  <lpage>-3491</lpage>
</bibl>

<bibl id="B43">
  <title><p>Gaussian processes for global optimization</p></title>
  <aug>
    <au><snm>Osborne</snm><fnm>MA</fnm></au>
    <au><snm>Garnett</snm><fnm>R</fnm></au>
    <au><snm>Roberts</snm><fnm>SJ</fnm></au>
  </aug>
  <source>in LION</source>
  <pubdate>2009</pubdate>
</bibl>

<bibl id="B44">
  <note>M. Abadi, {\it et al.}, TensorFlow: Large-scale machine learning on
  heterogeneous systems (2015). Software available from
  \url{tensorflow.org}.</note>
</bibl>

</refgrp>
} 

\end{backmatter}

\end{document}